# The Phase-1 Upgrade for the Level-1 Muon Barrel Trigger of the ATLAS Experiment at LHC

V. Izzo, A. Aloisio, R. Giordano, S. Perrella and R. Vari

*Abstract*–The Level-1 Muon Barrel Trigger of the ATLAS Experiment at LHC makes use of Resistive Plate Chamber (RPC) detectors. The on-detector trigger electronics modules are able to identify muons with predefined transverse momentum values ($p_T$) by executing a coincidence logic on signals coming from the various detector layers. Then, on-detector trigger boards transfer trigger data to the off-detector electronics. A complex trigger system processes the incoming data by combining trigger information from the Barrel and the End-cap regions, and by providing the combined muon candidate to the Central Trigger Processor (CTP).

For almost a decade, the Level-1 Trigger system has been operating very well, despite the challenging requirements on trigger efficiency and performance, and the continuously increasing LHC luminosity. In order to cope with these constraints, various upgrades for the full trigger system were already deployed, and others have been designed to be installed in the next years. Most of the upgrades to the trigger system rely on state-of-the-art technologies, thus allowing to increase the processing power and data transfer bandwidth, and to design more complex trigger menus. As a consequence, more trigger candidates might be selected by the system, thus supporting new physics studies or topological selections.

In this work, we discuss the design of the first prototype of the new Barrel Interface Board, designed around a Xilinx FPGA, which will be used to transfer RPC trigger data to the CTP system. The main requirement for the board is to support the optical transmission of the trigger data with fixed latency and to allow the implementation of new trigger algorithms. We describe the hardware implementation and the results of the first functional and integration tests.

## I. INTRODUCTION

THE Level-1 Muon Barrel trigger system [1] of the ATLAS detector [2] at CERN, is based on three concentric layers of Resistive Plate Chambers (RPC), which are used for muon detection (Fig.1). The on-detector electronics uses the hit signals on the RPCs, and triggers muons with specific values of transverse momentum ($p_T$); trigger electronics also tags the muon candidate tracks with the pertaining Bunch Crossing identifier (BCID). The trigger algorithm searches for hit coincidences between the various detector layers inside a programmed geometrical window, which is defined by the cut in $p_T$. The Barrel trigger is made of the low-$p_T$ and the high-$p_T$ algorithm, each using a specific selection scheme (Fig.2). Then, on-detector trigger boards transfer trigger data to the off-detector electronics in the USA15 counting room, via optical link. The Level-1 Muon Barrel in divided in 64 trigger sectors, according to the eightfold structure of the spectrometer. For each trigger sector, a Barrel Sector Logic (SL) board, installed in the USA15 counting room, receives the RPC trigger data, executes the trigger algorithm for its own sector, and produces a Muon Barrel trigger candidate, which is transferred to the Muon to CTP Interface (MuCTPI) board.

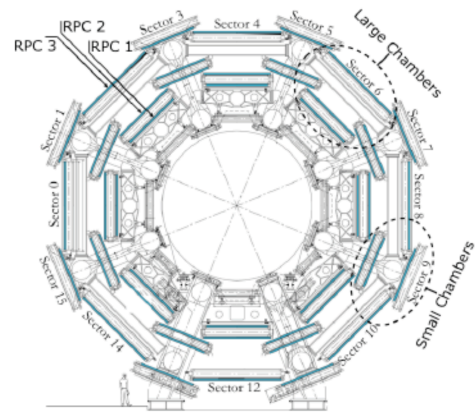

Fig. 1. The ATLAS Barrel region section on the azimuthal plane.[3]

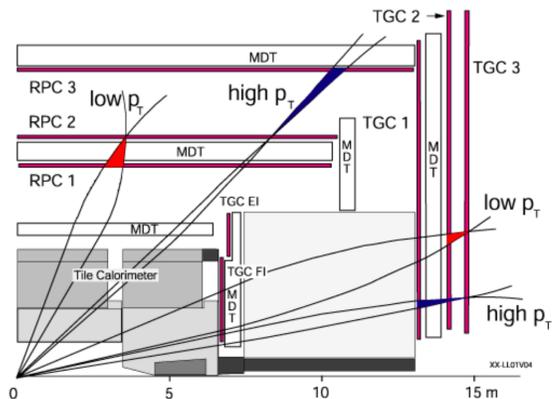

Fig. 2. The low-pT and high-pT trigger scheme, with the middle-pivot (RPC2), the inner low-pT (RPC1) and the outer high-pT (RPC3) detector. [4]

The MuCTPI board is responsible of receiving and processing the trigger data from the Barrel and the End-cap regions, thus providing the combined muon candidate to the Central Trigger Processor (CTP). The challenging requirement for the L1 trigger system is an event rate lowering, from a collision rate of 40 MHz to an overall acceptance rate of about 100 kHz, with a maximum fixed latency of 2.5μs.

Fig. 3 shows the muon trigger data flow from detectors to the CTP for the Barrel and End-cap region. An additional module is required for the Barrel system, namely the Barrel Interface Board acting as the interface between Barrel SL and MuCTPI.

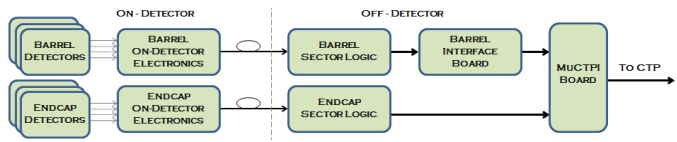

Fig. 3. The muon trigger data flow from the muon detectors to the CTP.

## II. MOTIVATION FOR THE UPGRADE

For almost a decade, the Level-1 Trigger system operated very well, despite the always-increasing challenges, concerning the requirements on trigger efficiency and performance, and the continuously increasing LHC luminosity. In order to cope with these constraints, various upgrades for the full trigger system were already deployed, and others have been designed to be installed in the next years. Most of the upgrades to the trigger system rely on state-of-the-art technologies, thus allowing to increase the processing power and data transfer bandwidth, and to design more complex trigger menus. As a consequence, more trigger candidates might be selected by the system, thus supporting new physics studies or topological selections.

According to the upgrade plans the MuCTPI board will be replaced for the Phase-I of the experiment [5]. The board will be fully redesigned and will be able to supply full-granularity muon information to L1 Calorimeter system, enabling the design of topological trigger algorithms using signals from both the calorimeter and muon systems. The MuCTPI will also be interfaced to the Muon SL boards via high-speed optical links. As a consequence, the data transfer bandwidth will be increased and new physics studies will be enabled by processing more than two trigger candidates per sector.

## III. IMPLEMENTATION

According to the ATLAS Phase-1 TDR [5], we designed a new version of the VME Barrel Interface Board which is able to transfer the Barrel trigger data to the new MuCTPI via an optical link. A Xilinx Artix-7 Field Programmable Gate Array device (FPGA) [6] equipped with high-speed embedded transceivers, is the main processor of the board. During the design of the new Barrel Interface Board we kept all the functionalities of the previous board; moreover, thanks to the large amount of resources available in the Artix-7 FPGA device, the new board might perform more complicated local trigger algorithms or include monitoring logic. It's worth to stress that the Barrel Interface Board will provide a much higher bandwidth (6.4 Gb/s) than the one available during the present runs of ATLAS (1.28 Gb/s). The bandwidth increase is possible by using the FPGA high-speed embedded transceivers, that will serialize and optically transmit Level-1 Barrel trigger data to the new MuCTPI. Moreover, the high-speed optical links will also fulfill the fixed latency requirement for trigger data.

## IV. TEST RESULTS

The first prototype of the Barrel Interface Board is fully working and was used in integration tests at CERN with the new MuCTPI board prototype, where the fixed latency transmission protocol at 6.4 Gbps was validated.

We paid a particular attention in verifying two main characteristics of the board: the SI5345 Jitter Cleaner functionalities and the 6.4 Gb/s data transfer capabilities.

Concerning the SI5345 Jitter Cleaner component, it produces a 320 MHz clock (to be used by the FPGA high-speed transceivers), with a constant phase relationship with the 40 MHz input (Zero Delay Mode). The Period Jitter on the output clock signal of the SI5345 was measured to be ~22 ps; the Cycle-to-Cycle jitter was measured to be ~38 ps.

The first prototype of the Barrel Interface Board was used in integration tests at CERN with the new MuCTPI board prototype: we characterized and validated the transmission protocol at 6.4 Gbps, 8B/10B encoded, with fixed latency. We used Xilinx IBERT tool to check the link stability with a BER (Bit Error Ratio) up to $10^{-15}$; we also measured the transmission latency to be less than 75 ns, which is definitely acceptable and below the limits defined by the experiment.

## V. CONCLUSIONS

We described the design of the first prototype of the new Barrel Interface Board designed around a Xilinx FPGA which will be used to transfer RPC trigger data to the MuCTPI board, for the Phase-I upgrade of the Level-1 Muon Barrel Trigger of the ATLAS experiment.

The board supports the optical transmission of the trigger data with fixed latency and might implement new trigger algorithms. We discussed the hardware implementation and the results of the first functional and integration tests at CERN.

All the features of the prototype of the Barrel Interface Board have been successfully tested; the production of the boards will start later 2018, and installation and commissioning will start in 2019.